\documentclass[11pt,letterpaper]{article}

\usepackage[utf8]{inputenc}
\usepackage[T1]{fontenc}
\usepackage{amsmath,amssymb,amsfonts}
\usepackage{booktabs}
\usepackage{caption}
\usepackage{float}
\usepackage{graphicx}
\usepackage{xcolor}
\usepackage{enumitem}
\usepackage{setspace}
\usepackage[margin=1in]{geometry}
\usepackage{fancyhdr}
\usepackage{titlesec}
\usepackage{abstract}
\usepackage{hanging}   
\usepackage{hyperref}
\hypersetup{
  colorlinks=false,
  pdfborder={0 0 0},
  pdfauthor={Ajay Kumar Verma, Jul Jon Ramirez General, Yvan Landry Ndzonde Fonkou}
}

\titleformat{\section}{\large\bfseries}{\thesection.}{0.5em}{}
\titleformat{\subsection}{\normalsize\bfseries}{\thesubsection}{0.5em}{}
\titleformat{\paragraph}[runin]{\normalsize\bfseries}{}{0em}{}[.]

\newenvironment{workscited}{%
  \section*{References}
  \begin{hangparas}{0.5in}{1}%
  \setlength{\parskip}{4pt}%
}{%
  \end{hangparas}%
}

\usepackage{xcolor} 
\usepackage{datetime}
\title{Deep Learning Forecasting of the U.S. Aggregate Bond Index}
\newdateformat{monthyear}{\monthname[\THEMONTH] \THEYEAR}

\author{
\textcolor{gray}{
Ajay Kumar Verma$^{1}$, 
Jul Jon Ramirez General$^{2}$, 
Yvan Landry Ndzonde Fonkou$^{3}$
}
}

\begin{document}

\date{\monthyear\today}

\maketitle

\renewcommand{\thefootnote}{}

\footnotetext{
\small
$^{1}$Independent Researcher (azay.verma@gmail.com) , 
$^{2}$Bangko Sentral ng Pilipinas (generaljr@bsp.gov.ph) , 
$^{3}$Independent Researcher (yvanndzonde@gmail.com)
}

\renewcommand{\thefootnote}{\arabic{footnote}}

\begin{abstract}
This study looks at the statistical properties and predictability using deep learning methods of the U.S.\ aggregate bond index in daily observations spanning
2018 to February 2026. We first establish that index \emph{levels} are extremely persistent
and consistent with unit-root behavior (Dickey and Fuller), while \emph{log
returns} are covariance-stationary with weak linear dependence and pronounced
volatility clustering characteristic of ARCH-type processes (Engle;
Bollerslev). Motivated by the trade-off between stationarity and information
retention, we construct a ``stationary but maximally persistent'' representation
via fractional differencing (Granger and Joyeux; Hosking) following the procedure of L\'{o}pez de Prado, and evaluate
short-horizon forecast using two neural paradigms: (i)~Multilayer Perceptrons
(MLPs) trained on lagged vectors with joint lag-length and hyperparameter tuning
(Hornik et al.; Rumelhart et al.); and (ii)~Convolutional Neural Networks (CNNs)
trained on Gramian Angular Field (GAF) image encodings (Wang
and Oates). Empirically, MLPs match the strong naive persistence benchmark on
levels, collapse toward near-zero forecasts on returns, and achieve the strongest
incremental performance on the fractionally differenced series, where moderate
dependence remains but unit-root drift is attenuated. In contrast, CNN--GAF
models deliver consistently negative out-of-sample $R^2$ across all three
representations. Overall, the results imply that, for
short-horizon forecasting of broad bond indices, the primary determinant of
predictive performance is the transformation of the series-its degree of
stationarity and memory-rather than architectural complexity. Lag-based models
remain competitive under persistence, while GAF-based CNNs are better suited to
pattern-based tasks than to persistence-dominated next-step prediction.

\end{abstract}

\newpage
\section{Introduction and Objectives}
\label{sec:intro}

The predictability of financial time series at short horizons
has been researched by financial economists and practitioners for decades (Fama; Campbell
and Shiller). The development of deep learning methods have renewed this inquiry, offering flexible function approximators capable of extracting nonlinear
structure from high-dimensional inputs (LeCun et al., ``Deep Learning''). But,
the application of such methods to financial forecasting considering the
statistical properties of index prices---non-stationarity, volatility clustering,
fat tails, etc - impose specific requirements on data preprocessing,
model design, and evaluation that differ substantially from those encountered in
image recognition or natural language processing (Hamilton).

The present study investigates the statistical properties and the short-horizon
predictability of the U.S.\ Aggregate Bond Index. Using 2{,}000 daily
observations spanning 2018--2026, we pursue seven interconnected objectives: First, to characterize the distributional behavior of index levels, including measures of dispersion, skewness, and kurtosis. Second, we formally assess
unit-root behavior in levels leveraging the Augmented Dickey--Fuller (ADF) test (Said
and Dickey), confirming whether the index is integrated of order one, $I(1)$.
Third, we establish that log returns are approximately $I(0)$ by documenting weak
autocorrelation, rapid mean reversion, and heteroskedastic variance dynamics.
Fourth, we implement the fractional differencing procedure of L\'{o}pez de Prado
to construct a near-stationary representation that preserves long-memory
structure, selecting the differencing order via an ADF $p$-value diagnostic.
Fifth, we tune the lag length and hyperparameters of a feed-forward MLP and test
whether nonlinear lag-based models improve one-step-ahead forecasts beyond naive
persistence benchmarks across levels, log returns and the fractionally differenced
series. Sixth, we transform each series into Gramian Angular Field images and check whether CNNs can extract predictive patterns from this image
representation. Finally, we compare the two architectures in terms of inductive
bias, representation choice, and the statistical properties of the underlying
time series, providing a theoretically grounded account of the observed
performance differences.

The study contributes to a growing literature on machine learning methods in
finance (L\'{o}pez de Prado; Gu et al.) by demonstrating that the selection of data
representation and not the model complexity is often the binding constraint on
predictive performance for highly persistent financial series.

\newpage
\section{Empirical Analysis of U.S.\ Aggregate Bond Index Levels}
\label{sec:levels}

\subsection{Sample and Distributional Properties}
\label{subsec:sample}

\begin{figure}[h!]
  \centering
  \includegraphics[width=\linewidth]{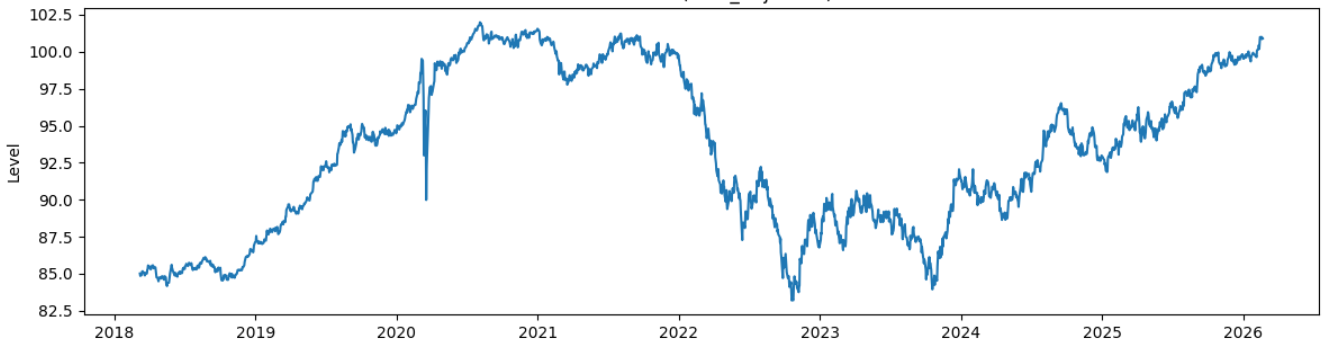}
  \caption{Daily closing levels of the U.S.\ Agg Bond Index,
           March 2018--February 2026.}
  \label{fig:level_ts}
\end{figure}

The sample spans from March 8, 2018 to February 20, 2026, comprising
$n = 2{,}000$ daily observations of index levels. Figure~\ref{fig:level_ts}
shows the full time series of the U.S.\ Agg Bond Index over this period.

Summary statistics for the index level are shown in
Table~\ref{tab:level_stats}. The unconditional distribution of the Index level
appears approximately symmetric: the estimated skewness of $-0.034$ is
economically negligible, and the excess kurtosis of $-1.298$ denotes a
platykurtic distribution relative to the Gaussian benchmark, with lighter tails
but a flatter peak. Such behavior is consistent with index levels observed over
a bounded sample horizon, where extreme realizations are naturally constrained
by the mean-reversion of long-horizon bond yields. Over the period, the index
fluctuates between $83.19$ and $101.99$, with a standard deviation of $5.34$.
The interquantile interval spanning the 5th to 95th percentiles (approximately
$85.0$ to $100.9$) confirms moderate unconditional dispersion. The maximum
drawdown of $-18.43\%$ represents the largest peak-to-trough reduction during the period and is consistent with the bond market changes associated with the
2022--2023 interest-rate normalization cycle (Hamilton).

Although the unconditional distribution appears stable in scale and symmetry,
price levels of broad fixed-income indices are typically highly persistent and
exhibit unit-root behavior (Nelson and Plosser). The relatively narrow dispersion
and bounded range therefore do not imply stationarity; formal statistical testing
is required. In financial economics, the standard paradigm distinguishes between
non-stationary levels ($I(1)$) and stationary first differences or log returns
($I(0)$), so that econometric inference is appropriately conducted on returns
rather than on raw price levels (Engle and Granger).

\begin{table}[h!]
\centering
\caption{Summary statistics for U.S.\ Aggregate Bond Index levels ($n=2{,}000$).}
\label{tab:level_stats}
\begin{tabular}{lr}
\toprule
Statistic & Value \\
\midrule
Mean & 93.260 \\
Standard Deviation & 5.335 \\
Minimum & 83.191 \\
1st Percentile & 84.372 \\
5th Percentile & 84.994 \\
Median & 93.600 \\
95th Percentile & 100.929 \\
99th Percentile & 101.398 \\
Maximum & 101.991 \\
Skewness & $-0.034$ \\
Excess Kurtosis & $-1.298$ \\
Maximum Drawdown & $-18.43\%$ \\
\bottomrule
\end{tabular}
\end{table}

\subsection{Autocorrelation Structure and Implications for Stationarity}
\label{subsec:acf_levels}

The autocorrelations of daily index levels are shown in
Table~\ref{tab:acf_levels}. The lag-1 auto-correlation is exceptionally high
($\rho_1 = 0.998$), indicating near-perfect persistence between consecutive
trading days. Even at a monthly horizon of 21 trading days, the autocorrelation
remains strong ($\rho_{21} = 0.963$), and at the quarterly frequency
($\rho_{63} = 0.874$) it remains substantial. Only at the annual horizon does
persistence reduce more noticeably ($\rho_{252} = 0.265$), yet it remains
clearly positive.

\begin{table}[h!]
\centering
\caption{Sample autocorrelations of U.S.\ Aggregate Bond Index levels at
         selected lags.}
\label{tab:acf_levels}
\begin{tabular}{lr}
\toprule
Lag & Autocorrelation \\
\midrule
1 day & 0.998 \\
5 days & 0.991 \\
21 days (1 month) & 0.963 \\
63 days (1 quarter) & 0.874 \\
252 days (1 year) & 0.265 \\
\bottomrule
\end{tabular}
\end{table}

This slowly decaying autocorrelation function is a hallmark of integrated
processes (Box et al.). For a covariance-stationary process, autocorrelations
generally decline toward zero at a geometric rate, whereas unit-root processes exhibit very high autocorrelations at short lags and only gradual mean
reversion. The observed pattern is consistent with the behavior of an $I(1)$
process of below form
\begin{equation}
P_t = P_{t-1} + \varepsilon_t,
\label{eq:rw}
\end{equation}
where $\varepsilon_t$ is a mean-zero innovation. This provides preliminary
evidence against stationarity in levels, which motivates formal unit-root testing and the subsequent analysis of log returns and fractionally differenced
transformations.

\subsection{Unit-Root Testing using the Augmented Dickey--Fuller Test}
\label{subsec:adf_levels}

To formally evaluate the stationarity of the index level, we use the
Augmented Dickey--Fuller (ADF) test (Dickey and Fuller; Said and Dickey) under
two specifications: a) a model with a constant only, and b) a model with both a
constant plus a deterministic time trend. The ADF test evaluates the null
hypothesis ($H_0$) that the series contains a unit root.

\begin{table}[h!]
\centering
\caption{ADF test for U.S.\ Aggregate Bond Index levels (constant only).}
\label{tab:adf_const}
\begin{tabular}{lr}
\toprule
Statistic & Value \\
\midrule
ADF Statistic & $-1.3715$ \\
$p$-value & 0.5959 \\
Lags Used & 7 \\
Observations & 1{,}992 \\
Critical Value (1\%) & $-3.4336$ \\
Critical Value (5\%) & $-2.8630$ \\
Critical Value (10\%) & $-2.5675$ \\
AIC & 1{,}328.415 \\
\bottomrule
\end{tabular}
\end{table}

\begin{table}[h!]
\centering
\caption{ADF test for U.S.\ Aggregate Bond Index levels (constant and trend).}
\label{tab:adf_trend}
\begin{tabular}{lr}
\toprule
Statistic & Value \\
\midrule
ADF Statistic & $-1.3798$ \\
$p$-value & 0.8667 \\
Lags Used & 7 \\
Observations & 1{,}992 \\
Critical Value (1\%) & $-3.9633$ \\
Critical Value (5\%) & $-3.4127$ \\
Critical Value (10\%) & $-3.1283$ \\
AIC & 1{,}330.387 \\
\bottomrule
\end{tabular}
\end{table}

Under the constant-only specification, the test statistic of $-1.3715$ is
substantially above all critical values in absolute magnitude
($-1.3715 > -2.8630$ at the 5\% level), and the $p$-value of $0.596$ is far
above conventional significance thresholds, thus providing no evidence against the
null hypothesis of having a unit root. Adding a deterministic time trend does not alter
this conclusion: the test statistic of $-1.3798$ remains above critical
values, with the $p$-value rising to $0.867$, yielding even weaker evidence
against $H_0$. Taken together with the extremely high short-horizon
autocorrelations and the slow decay of the empirical autocorrelation function
documented above, both ADF specifications strongly support the characterizing the U.S.\ Aggregate Bond Index level series as an integrated process of order
one, $I(1)$. Hence, statistical inference, volatility modeling, and
predictive analysis should be conducted on first differences (log returns) rather
than on raw levels.

\section{Log Returns: Stationarity, Dependence, and Volatility Dynamics}
\label{sec:returns}

\subsection{Visual Inspection and Mean Behavior}
\label{subsec:returns_visual}

\begin{figure}[h!]
  \centering
  \includegraphics[width=\linewidth]{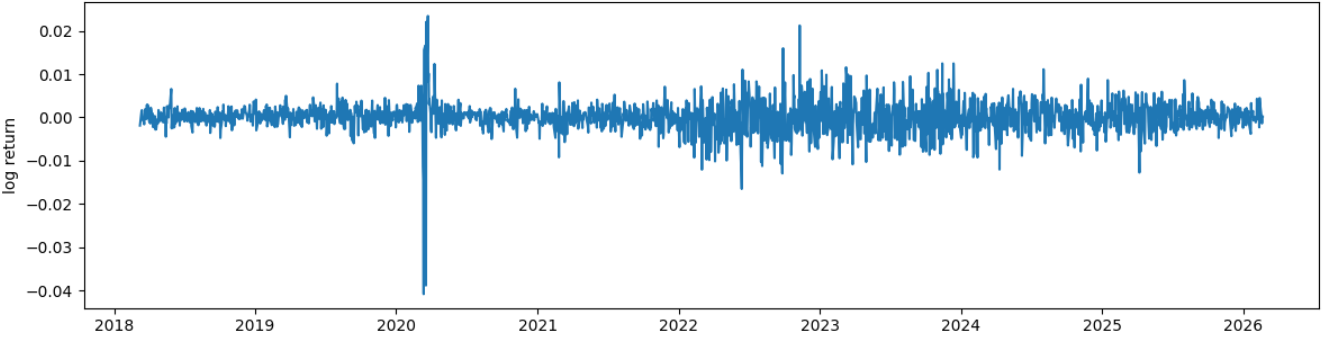}
  \caption{Daily log returns of the U.S.\ Aggregate Bond Index, 2018--2026.}
  \label{fig:returns_ts}
\end{figure}

Figure~\ref{fig:returns_ts} shows the daily log returns of the U.S.\
Aggregate Bond Index. Log returns correspond to first differencing in
logarithms. If the level process is approximately $I(1)$ as established in
Section~\ref{sec:levels}, the log-return series should be approximately
$I(0)$ stationary with finite mean and variance (Nelson and Plosser). The
series hovers around zero with no visible deterministic trend, strongly
supporting this characterization.

A key feature visible in the figure is volatility clustering: distinct periods
of low volatility (2018--2019) proceeds with a sharp spike in early 2020
associated with the COVID-19-related bond market shock, elevated and persistent
volatility during 2022--2023 corresponding to the interest-rate normalization
cycle, and gradual moderation thereafter. This pattern is a canonical stylized
fact of financial time series (Cont), commonly modeled using ARCH and GARCH
frameworks (Engle; Bollerslev), and suggests that conditional heteroskedasticity
is a dominant feature of this series. The occasional large spikes in both
directions imply heavy-tailed
return distributions relative to the Gaussian benchmark and potential asymmetry
during stress episodes, consistent with the broader literature on fat tails in
financial returns (Mandelbrot).

\subsection{Formal Stationarity Tests and Autocorrelation Structure}
\label{subsec:returns_adf}

Table~\ref{tab:adf_returns} shows the results of the ADF test for the log-return
series. The test statistic of $-17.8123$ is below all critical values in
absolute magnitude ($-17.8123 < -3.4336$ at the 1\% level), and the $p$-value
is effectively zero, leading to decisive rejection of the unit-root null
hypothesis. This provides overwhelming statistical evidence that the log-return
series is covariance-stationary.

\begin{table}[h!]
\centering
\caption{ADF test for U.S.\ Aggregate Bond Index log returns (constant
         specification).}
\label{tab:adf_returns}
\begin{tabular}{lr}
\toprule
Statistic & Value \\
\midrule
ADF Statistic & $-17.8123$ \\
$p$-value & $< 0.001$ \\
Lags Used & 6 \\
Observations & 1{,}992 \\
Critical Value (1\%) & $-3.4336$ \\
Critical Value (5\%) & $-2.8630$ \\
Critical Value (10\%) & $-2.5675$ \\
AIC & $-16{,}504.110$ \\
\bottomrule
\end{tabular}
\end{table}

The autocorrelation structure of the log-return series, shown in
Table~\ref{tab:acf_returns}, is in stark contrast to the level series.
Autocorrelations are small in magnitude over all lags: the lag-1
autocorrelation is $0.027$, and at the quarterly horizon ($63$ trading days)
it remains modest at $0.064$; at the annual horizon it is essentially zero.
This rapid decay toward zero is characteristic of stationary return processes
and aligns with the weak-form efficiency hypothesis (Fama), under which
linear structure predictability in returns is limited. Formally, the evidence
supports the representation
\begin{equation}
r_t = \mu + \varepsilon_t,
\label{eq:returns}
\end{equation}
where $\varepsilon_t$ is a mean-zero stationary disturbance exhibiting
conditional heteroskedasticity. These findings establish that the U.S.\
Aggregate Bond Index is well-characterized as an $I(1)$ process in levels and
an $I(0)$ process in returns, consistent with canonical asset pricing theory and
the standard prescription for modeling and forecasting (Campbell and Shiller).

\begin{table}[h!]
\centering
\caption{Sample autocorrelations of U.S.\ Aggregate Bond Index log returns.}
\label{tab:acf_returns}
\begin{tabular}{lr}
\toprule
Lag & Autocorrelation \\
\midrule
1 day & 0.027 \\
5 days & 0.002 \\
21 days (1 month) & 0.019 \\
63 days (1 quarter) & 0.064 \\
252 days (1 year) & $-0.002$ \\
\bottomrule
\end{tabular}
\end{table}

\newpage
\section{Fractional Differencing: Theory, Diagnostics, and Implementation}
\label{sec:fracdiff}

\subsection{Classical Differencing and the Over-Differencing Problem}
\label{subsec:overdiff}

The empirical results of Sections~\ref{sec:levels} and~\ref{sec:returns}
show that index levels are $I(1)$ and that log returns are $I(0)$, which
justify first differencing (in logarithms) as the canonical transformation for
reducing the stochastic trend. An important concern in this context is whether
first differencing is excessive, that is, whether the original series were
already stationary and differencing merely introduces artificial negative
autocorrelation and amplifies noise. The empirical autocorrelation structure of
returns does not exhibit such behavior, and the ADF test for levels provides no
evidence of prior stationarity; first differencing is therefore appropriate (Box
et al.).

However, integer differencing can also be problematic in a different direction:
by applying the operator $(1-L)^1$ to eliminate the stochastic trend, one may
inadvertently discard low-frequency dependence that carries economically relevant
predictive information. This observation motivates the more general framework of
fractional differencing (Granger and Joyeux; Hosking), in which the differencing
operator is parameterized continuously as $(1-L)^d$ for $d \in (0,1)$. Using the
binomial expansion,
\begin{equation}
(1-L)^d = \sum_{k=0}^{\infty} w_k L^k,
\qquad
w_k = (-1)^k \binom{d}{k}
    = (-1)^k \frac{d(d-1)\cdots(d-k+1)}{k!},
\label{eq:fracdiff_op}
\end{equation}
the fractionally differenced series is
\begin{equation}
\tilde{P}_t = (1-L)^d P_t = \sum_{k=0}^{\infty} w_k P_{t-k}.
\label{eq:fracdiff_series}
\end{equation}
The weights $w_k$ decay hyperbolically at rate $|w_k| \sim k^{-(1+d)}$ for large
$k$, generating the long-memory dependence that distinguishes fractional
differencing from integer differencing (where all weights beyond $k=1$ are
exactly zero for $d=1$). Unlike exponential decay, this hyperbolic decay
persists low-frequency structure in the transformed series. The practical
significance for machine learning pipelines is threefold: many algorithms assume
stationarity implicitly; excessive differencing can degrade the signal-to-noise
ratio; and long-memory components may contain predictive features that would be
destroyed by full integer differencing (Baillie; L\'{o}pez de Prado).
\newpage
\subsection{The L\'{o}pez de Prado ``Stationary but Maximally Persistent''
  Criterion}
\label{subsec:ldp}

L\'{o}pez de Prado proposes selecting the minimum differencing order $d^*$ such
that the transformed series achieves stationarity, while retaining the maximum
admissible long-memory structure:
\begin{equation}
d^* = \min \{ d \mid (1-L)^d P_t \text{ is stationary} \}.
\label{eq:dstar}
\end{equation}
Generally, we evaluate the fractional differencing transformation over a grid
of $d$ values, apply an ADF test to each transformed series, and select the
smallest $d$ for which the unit-root hypothesis is rejected at a chosen
significance level. The philosophical distinction from classical econometrics is
important: the standard approach removes integration fully by setting
$d=1$, while the fractional ML approach eliminates only the minimum integration
required, preserving the maximum amount of memory for downstream modeling. This
is particularly relevant in a machine learning context, where long-memory
features may be informative predictors that would be lost under full differencing
(Baillie; L\'{o}pez de Prado).

\begin{figure}[h!]
  \centering
  \includegraphics[width=\linewidth]{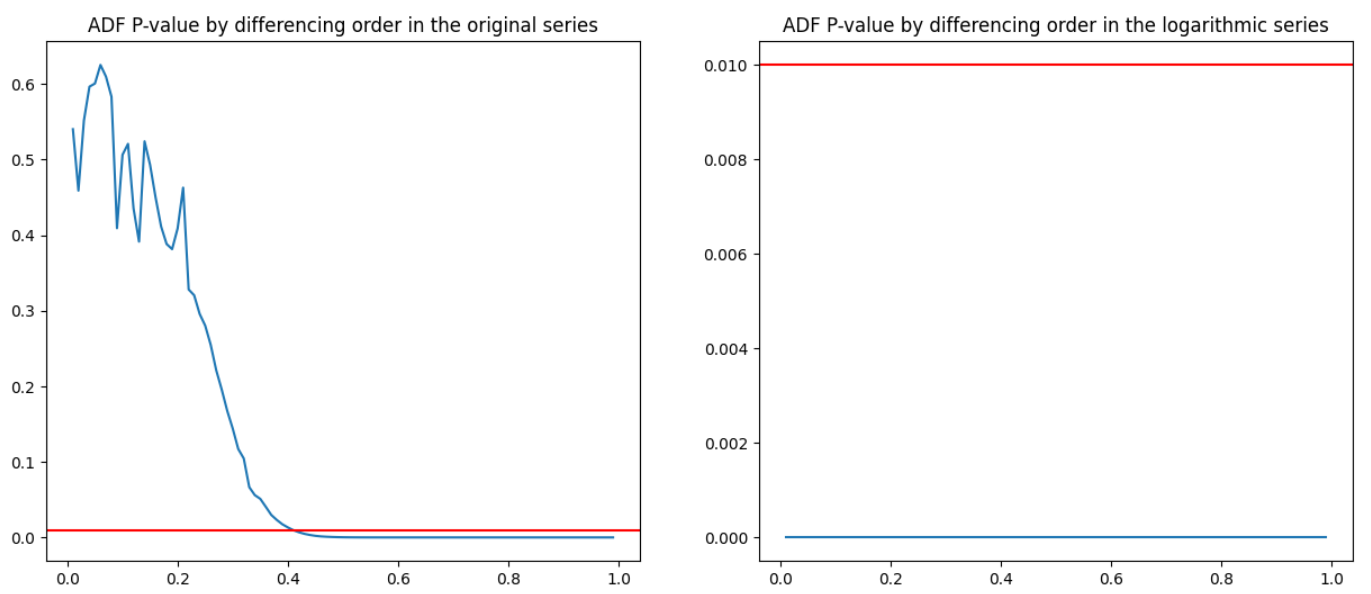}
  \caption{ADF $p$-value as a function of differencing order $d$ for the
           original-scale level series (left) and the log-transformed series
           (right). The red horizontal line denotes the 1\% significance
           threshold.}
  \label{fig:fracdiff_adf}
\end{figure}

Figure~\ref{fig:fracdiff_adf} illustrates this diagnostic for the U.S.\ Aggregate
Bond Index. The left panel reports ADF $p$-values as a function of
$d \in [0,1]$ for the original-scale level series, while the right panel shows
the corresponding diagnostics for the log transformed series.

For the original levels, ADF $p$-values remain above conventional
significance thresholds for small $d$, confirming the failure to reject the
unit-root null hypothesis documented in Section~\ref{sec:levels}. As $d$ increases, the
$p$-value declines monotonically, crossing the 1\% threshold only at
approximately $d^* \approx 0.40$--$0.45$. This implies that a meaningful degree
of differencing is required before the series achieves stationarity under the ADF
tests, and that fractional differencing at $d \approx 0.4$ provides a
near-stationary representation while retaining substantially more memory than
full integer differencing. For the log-transformed series, the ADF $p$-value is
effectively zero across the entire grid, confirming that log returns are
unambiguously stationary for any positive differencing order.

\subsection{The Fractionally Differenced Level Series at $d=0.4$}
\label{subsec:fracdiff_d04}

\begin{figure}[h!]
  \centering
  \includegraphics[width=\linewidth]{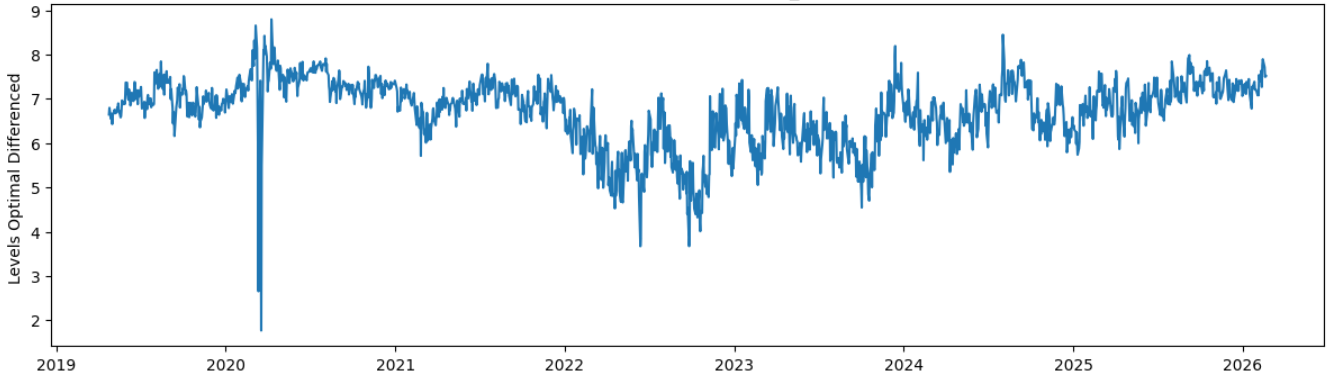}
  \caption{U.S.\ Aggregate Bond Index levels after fractional differencing at
           $d=0.4$ (``stationary but maximally persistent'' representation).}
  \label{fig:fracdiff_ts}
\end{figure}

Motivated by the diagnostic of Figure~\ref{fig:fracdiff_adf}, we apply
fractional differencing at $d=0.4$ to the U.S.\ Aggregate level series, arriving at
the transformed process
\begin{equation}
\tilde{P}^{(0.4)}_t = (1-L)^{0.4} P_t = \sum_{k=0}^{K} w_k P_{t-k},
\label{eq:fracdiff04}
\end{equation}
when the infinite expansion is truncated at a finite lag $K$ using a
weight-threshold criterion ($|w_K| < 10^{-5}$). Because the operator requires
$K$ lagged observations, the effective sample spans from April 25, 2019 to
February 20, 2026, with $n=1{,}716$ observations. The transformed series is
shown in Figure~\ref{fig:fracdiff_ts}.

Several features of the transformed series warrant discussion. The dominant
stochastic trend present in the raw level series is materially attenuated: the
series fluctuates around a relatively stable band, consistent with
near-stationarity. At the same time, slow-moving multi-month swings persist in
the transformed series, reflecting the deliberate preservation of low-frequency
dependence relative to full first differencing (Hosking). A pronounced negative
spike in early 2020 reflects the response of the fractional filter which
applies a weighted sum of many past observations to the sharp COVID-19 shock
in the underlying level series. Such spikes are informative about structural
breaks and market dislocations but can also create influential outliers in model
estimation, warranting robust scaling or outlier-aware loss functions. Elevated
variability during 2022--2023 is consistent with the interest-rate regime shift
and suggests that conditional heteroskedasticity persists even after fractional
differencing.

Summary statistics and autocorrelations for the transformed series are presented
in Tables~\ref{tab:fracdiff_stats} and~\ref{tab:fracdiff_acf}, respectively. The
distribution is asymmetric (skewness $= -1.09$) and heavy-tailed (excess kurtosis
$= 2.22$), reflecting the influence of stress episodes that manifest as
transitory shocks after applying the long-memory filter. The lag-1
autocorrelation reduces from near unity in levels to $0.860$ after fractional
differencing, with smooth decay across horizons precisely the intended
intermediate profile between the near-random-walk behavior of levels and the
near-white-noise behavior of returns. Formal ADF testing (Table~\ref{tab:fracdiff_adf})
under the constant-only specification yields a $p$-value of $0.013$, marginally
rejecting the unit-root null at the 5\% level; under the constant-plus-trend
specification, the $p$-value rises to $0.062$, reflecting borderline behavior
consistent with the deliberate choice of a small $d$ to preserve memory. This
borderline stationarity is not a defect of the transformation; it reflects the
intended trade-off between integration removal and memory preservation.

\begin{table}[h!]
\centering
\caption{Summary statistics for fractionally differenced levels ($d=0.4$,
         $n=1{,}716$).}
\label{tab:fracdiff_stats}
\begin{tabular}{lr}
\toprule
Statistic & Value \\
\midrule
Mean & 6.716 \\
Standard Deviation & 0.743 \\
Minimum & 1.768 \\
1st Percentile & 4.558 \\
5th Percentile & 5.300 \\
Median & 6.869 \\
95th Percentile & 7.648 \\
99th Percentile & 7.951 \\
Maximum & 8.796 \\
Skewness & $-1.088$ \\
Excess Kurtosis & 2.215 \\
\bottomrule
\end{tabular}
\end{table}

\begin{table}[h!]
\centering
\caption{Sample autocorrelations of fractionally differenced levels ($d=0.4$).}
\label{tab:fracdiff_acf}
\begin{tabular}{lr}
\toprule
Lag & Autocorrelation \\
\midrule
1 day & 0.860 \\
5 days & 0.713 \\
21 days (1 month) & 0.565 \\
63 days (1 quarter) & 0.398 \\
252 days (1 year) & 0.189 \\
\bottomrule
\end{tabular}
\end{table}

\begin{table}[h!]
\centering
\caption{ADF tests for fractionally differenced levels ($d=0.4$).}
\label{tab:fracdiff_adf}
\begin{tabular}{lrrrr}
\toprule
Specification & ADF Stat.\ & $p$-value & 5\% Critical & Decision (5\%) \\
\midrule
Constant ($c$) & $-3.338$ & 0.013 & $-2.863$ & Reject $H_0$ \\
Constant + Trend ($ct$) & $-3.327$ & 0.062 & $-3.413$ & Fail to reject $H_0$ \\
\bottomrule
\end{tabular}
\end{table}

The $d=0.4$ transformation can be interpreted as a principled compromise between
raw levels and log returns. Relative to levels ($d=0$), it is substantially
closer to stationary and more suitable for machine learning algorithms that
assume distributional stability over time (L\'{o}pez de Prado). Relative to log
returns ($d=1$), it retains richer low-frequency dependence that may be exploited
by predictive models if those dependencies are genuine rather than attributable
to stochastic drift (Baillie). In practice, the utility of the $d=0.4$
representation depends on whether the retained persistence is economically
meaningful - a question addressed empirically in the forecasting experiments of
Sections~\ref{sec:mlp} and~\ref{sec:cnn}.

\section{MLP Forecasting with Joint Lag and Hyperparameter Tuning}
\label{sec:mlp}

\subsection{Problem Formulation and Supervised Learning Setup}
\label{subsec:mlp_formulation}

We frame one-step-ahead forecasting as a supervised regression problem by
converting the univariate series $\{y_t\}_{t=1}^T$ into a design matrix via lag
embedding. For a candidate lag length $L \in \mathcal{L}$, the feature vector is
\begin{equation}
X_t = \bigl[y_{t-1},\, y_{t-2},\, \ldots,\, y_{t-L}\bigr]^\top \in \mathbb{R}^{L},
\qquad \text{with response } y_t,
\label{eq:lag_embed}
\end{equation}
for all $t$ such that the $L$ required lags exist, yielding a design matrix
$X \in \mathbb{R}^{N \times L}$ and response vector $y \in \mathbb{R}^{N}$ with
$N = T-L$. The forecast target is strictly one-step-ahead and uses only lagged
realizations, with no contemporaneous or future information incorporated into the
feature set - a requirement that is essential for causal validity in time-series
forecasting (Box et al.).

To preserve the chronological ordering of the data and avoid look-ahead bias, we
employ a strict temporal train--test split with test fraction $\alpha = 0.20$.
The first $(1-\alpha)N$ observations form the training set and the remaining
$\alpha N$ form the test set. Feature standardization uses a scaler fitted
exclusively on the training data, with training-derived mean and standard
deviation applied to the test set:
\begin{equation}
\tilde{X}_{\text{train}} = \text{Std}(X_{\text{train}}),
\qquad
\tilde{X}_{\text{test}} = \text{Std}_{\text{train}}(X_{\text{test}}),
\end{equation}
so that the test set exerts no influence on the scaling parameters. As a
benchmark, we employ the naive last-value forecast
$\hat{y}^{\text{naive}}_t = y_{t-1}$, which is known to be difficult to beat for
persistent processes (Meese and Rogoff) and serves as a strong practical
reference.

\subsection{MLP Architecture and Hyperparameter Search}
\label{subsec:mlp_arch}

The MLP maps $\mathbb{R}^{L} \to \mathbb{R}$ through a stack of fully-connected
layers:
\begin{equation}
f_\theta(X_t)
= W_K \phi(\cdots \phi(W_2 \phi(W_1 X_t + b_1) + b_2) \cdots) + b_K,
\end{equation}
where $\phi(\cdot)$ is a ReLU nonlinearity (Nair and Hinton). The universal
approximation theorem (Hornik et al.) guarantees that sufficiently deep or wide
MLPs can represent any continuous function to arbitrary accuracy, providing
theoretical justification for the architecture. Hyperparameters - number of
hidden layers ($K \in \{1,\ldots,6\}$), units per hidden layer
($\{16, 32, \ldots, 256\}$), dropout rate (Srivastava et al.)
($\{0.0, 0.1, 0.2, 0.3\}$), batch normalization (Ioffe and Szegedy) (on/off),
and Adam (Kingma and Ba) learning rate (log-uniform on
$[10^{-4}, 5\times10^{-3}]$) are selected using the Keras Tuner Hyperband
algorithm. All models minimize mean absolute error (MAE) on the training set
with an internal 20\% validation split; early stopping restores the best weights
and reduces overfitting.

Lag length $L$ changes the input dimensionality and must be tuned in an explicit
outer loop over a candidate grid $\mathcal{L}$. For each candidate $L$, we
construct the lagged dataset, apply the chronological split, run a Hyperband
search to identify the best MLP configuration, and record the best validation
loss. The lag length $L^*$ achieving the minimum validation loss is selected for
the final model. After identifying $L^*$ and the associated architecture, batch
size is further refined over $\{16, 32, 64, 128\}$ by selecting the value
yielding the lowest validation loss under early stopping. Out-of-sample
performance is shown for both the tuned MLP and the naive benchmark using
MAE, RMSE, $R^2$, and the Pearson correlation between predictions and realized
values.

\subsection{Empirical Results}
\label{subsec:mlp_results}

\begin{figure}[h!]
  \centering
  \includegraphics[width=\linewidth]{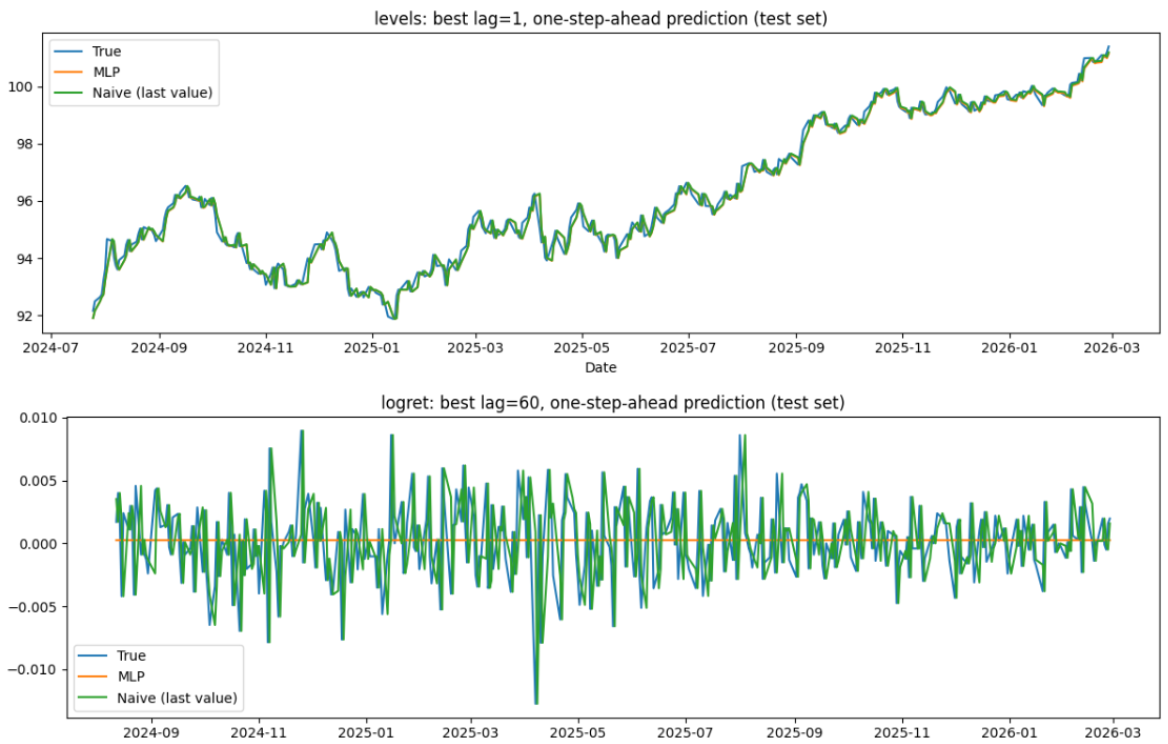}
  \caption{MLP and naive benchmark forecast trajectories for levels and log
           returns on the test segment.}
  \label{fig:mlp1}
\end{figure}

\begin{figure}[h!]
  \centering
  \includegraphics[width=\linewidth]{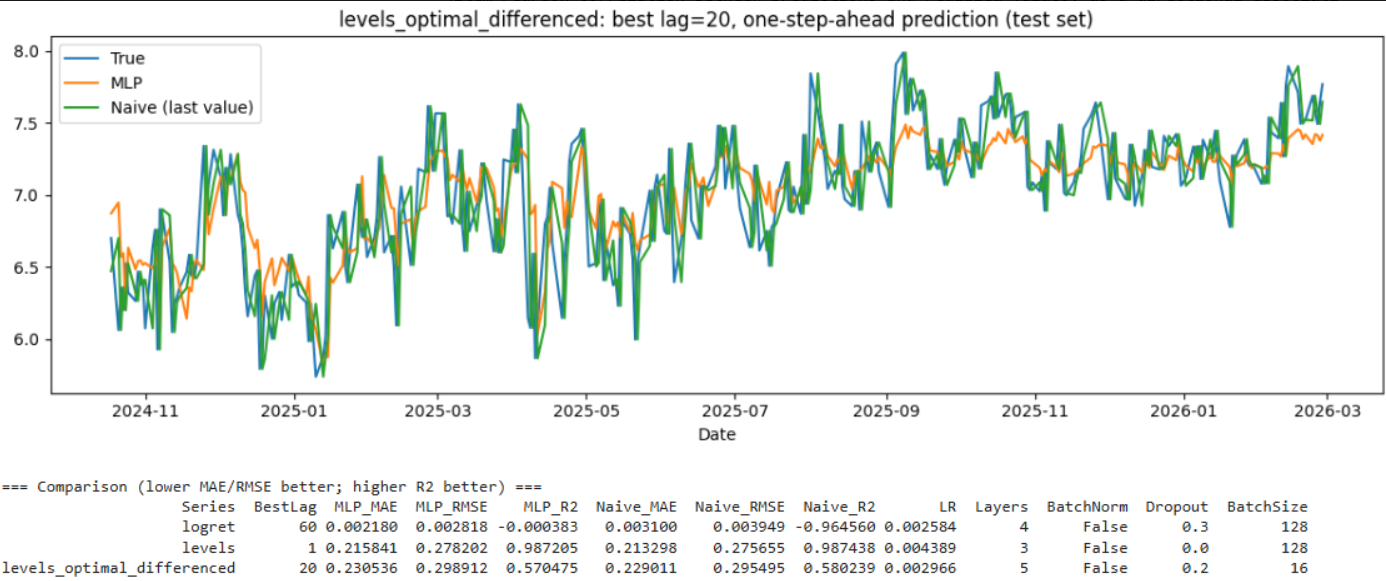}
  \caption{MLP and naive benchmark forecast trajectories for the fractionally
           differenced series ($d=0.4$) on the test segment.}
  \label{fig:mlp2}
\end{figure}

The forecasting framework was applied independently to the three representations
of the same underlying process: raw levels, log returns, and the fractionally
differenced series ($d=0.4$). Figures~\ref{fig:mlp1} and~\ref{fig:mlp2} shows
the forecast trajectories overlaid on the realized test-set observations for each
series.

\paragraph{Raw levels.}
The tuning procedure selected a lag length of $L=1$, consistent with the near
unit-root dynamics documented in Section~\ref{sec:levels}. With only the
immediately preceding observation as input, the MLP learns a near-identity
mapping, and its forecast closely overlaps with the naive benchmark throughout
the test period. Both achieve near-identical error magnitudes (MAE $\approx 0.216$
for the MLP versus $\approx 0.213$ for the naive forecast) and
$R^2 \approx 0.987$, indicating that almost all predictable variation arises from
simple persistence rather than nonlinear structure. This outcome is theoretically
expected: when the underlying process is nearly random walk, the conditional
expectation $\mathbb{E}[y_t \mid y_{t-1}]$ is well approximated by the identity
$y_{t-1}$, leaving little room for additional modeling capacity (Campbell and
Shiller).

\paragraph{Log returns.}
The return series exhibits substantially weaker serial dependence, and the tuning
procedure selected a longer memory window ($L=60$), reflecting the possibility of
weak predictive signals at longer horizons. Despite this, the resulting MLP
forecast is essentially flat around zero throughout the test sample, while
realized returns fluctuate with high volatility. The MLP achieves
MAE $\approx 0.00218$ - marginally better than the naive benchmark
(MAE $\approx 0.00310$) - and an $R^2$ of approximately $-0.00038$, indicating
performance no better than the unconditional mean predictor. These results are
in line with the efficient market hypothesis under weak-form informational
efficiency (Fama): past return information contains minimal predictive content
for next-day returns, and the neural network effectively learns that the optimal
forecast is close to zero.

\paragraph{Fractionally differenced levels.}
The optimally differenced series presents an intermediate case. The lag-selection
procedure identified $L=20$, and the MLP achieves MAE $\approx 0.231$ and
RMSE $\approx 0.299$, with $R^2 \approx 0.57$, indicating that a good
portion of the predictable variation in the transformed process is captured.
Unlike the level series, where the MLP and naive benchmark are essentially
identical, the MLP outperforms the naive forecast in certain local segments of
the test set and provides smoother tracking of the series' broader dynamics.
This result is consistent with the design objective of the fractional differencing
transformation: retaining moderate temporal dependence that nonlinear models can
exploit, while removing the dominant stochastic trend that renders the naive
forecast nearly optimal for raw levels.

\paragraph{Cross-series synthesis.}
The empirical pattern across the three transformations reveals a structural
principle: the effectiveness of machine learning models depends not on
architectural expressiveness alone, but critically on the signal-to-noise
properties of the input representation (Gu et al.). Raw levels are dominated by
persistence, for which the naive last-value forecast is nearly Bayes-optimal; log
returns exhibit minimal predictable linear structure, causing the network to
converge toward a near-zero forecast; the fractionally differenced representation
offers the most favorable trade-off, where moderate dependence survives but
extreme persistence has been attenuated. These findings reinforce the view that
appropriate preprocessing and stationarity-inducing transformations play a central
role in extracting predictive signals from financial time series (L\'{o}pez de
Prado).

\section{CNN Forecasting via Gramian Angular Field Representations}
\label{sec:cnn}

\subsection{Converting Time Series to Images via Gramian Angular Fields}
\label{subsec:gaf}

Convolutional neural networks (LeCun et al., ``Backpropagation'') require
structured two-dimensional inputs and are not directly applicable to raw
numerical time-series vectors. To bridge this gap, we employ the Gramian Angular
Field (GAF) methodology of Wang and Oates, which encodes a univariate
time-series window into a two-dimensional image that preserves temporal
dependencies and structural patterns accessible to convolutional feature
detectors. The GAF construction occurs in three steps. First, the time series
within a rolling window of length $W$ is rescaled to the interval $[-1, 1]$ via
min-max normalization. Second, each rescaled observation
$\tilde{y}_{t-k} \in [-1,1]$ is mapped to an angular value in polar coordinates
via
\begin{equation}
\phi_{t-k} = \arccos(\tilde{y}_{t-k}), \qquad k = 0, 1, \ldots, W-1.
\label{eq:gaf_angle}
\end{equation}
Third, a Gram matrix is constructed using the cosine of the pairwise sum of
angles:
\begin{equation}
G_t(i,j) = \cos(\phi_{t-i} + \phi_{t-j}),
\qquad i,j \in \{0, 1, \ldots, W-1\},
\label{eq:gaf_matrix}
\end{equation}
giving a symmetric $W \times W$ image that encodes pairwise temporal
relationships across the window. The diagonal entries recover the original
rescaled series, while off-diagonal entries encode inter-temporal correlations.
The GAF representation has been shown to be effective for time-series
classification tasks where recurring patterns, motifs, or regime signatures are
discriminative features (Wang and Oates; Fawaz et al.).

Figures below show the GAF encoding process and resulting
image representations for the level, log-return, and fractionally differenced
series.

\begin{figure}[H]
  \centering
  \includegraphics[width=\linewidth]{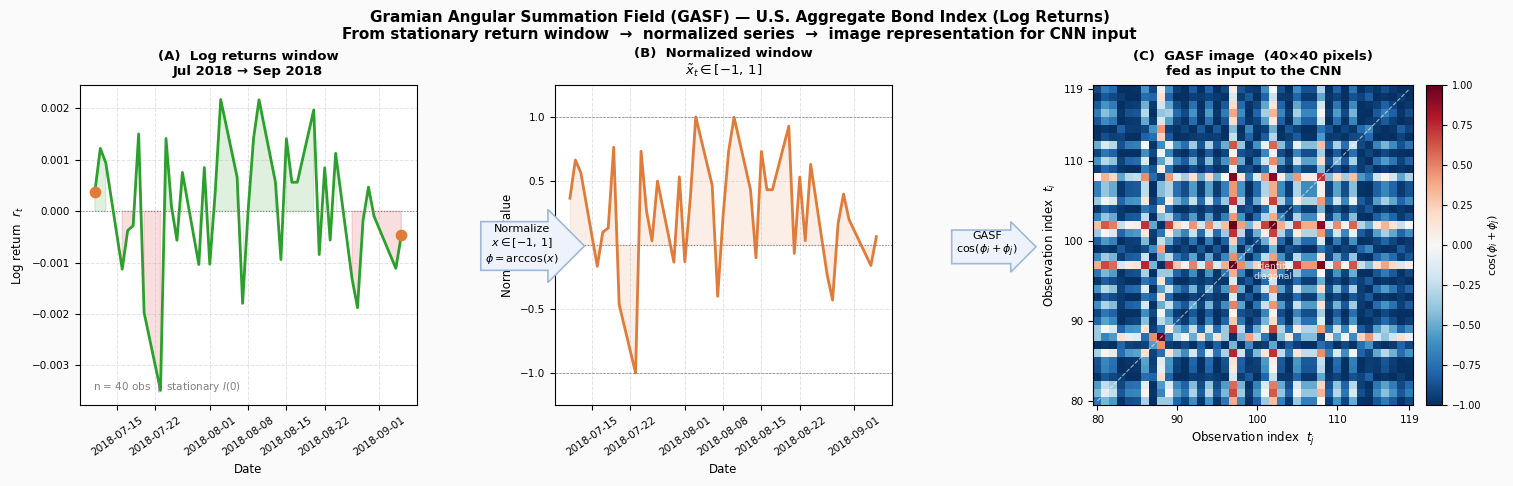}
  \caption{GAF encoding pipeline and the corresponding image representation for the
           level series.}
  \label{fig:gaf_levels}
\end{figure}

\begin{figure}[H]
  \centering
  \includegraphics[width=\linewidth]{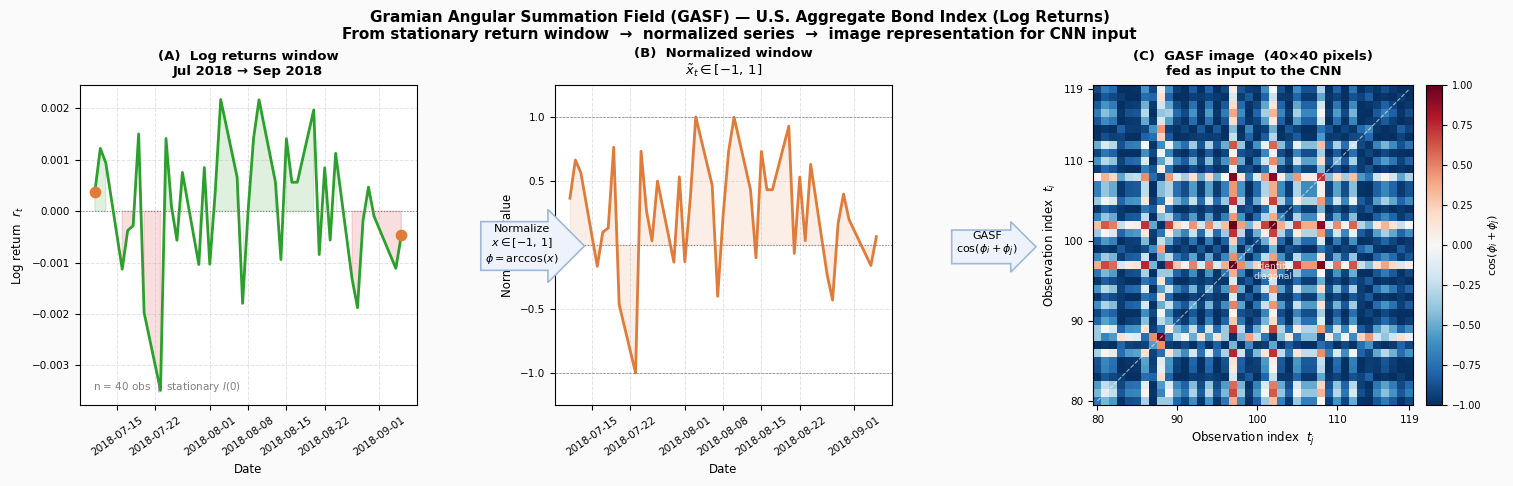}
  \caption{GAF encoding pipeline and the corresponding image representation for the
           log-return series.}
  \label{fig:gaf_returns}
\end{figure}

\begin{figure}[H]
  \centering
  \includegraphics[width=\linewidth]{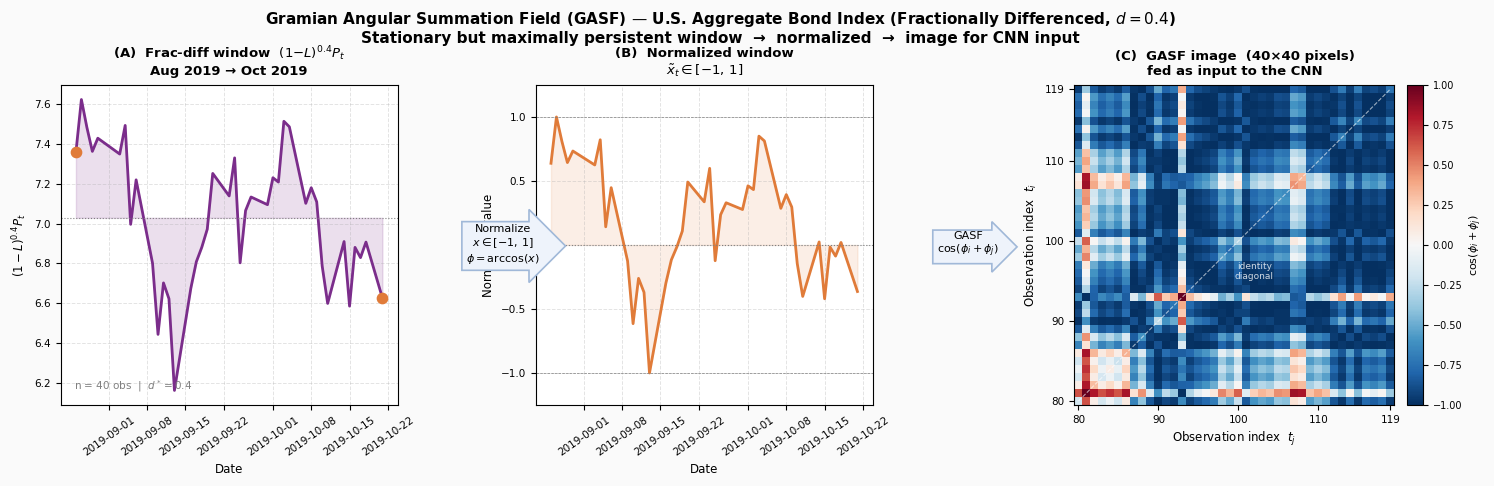}
  \caption{GAF encoding pipeline and the corresponding image representation for the
           fractionally differenced series ($d=0.4$).}
  \label{fig:gaf_fracdiff}
\end{figure}

\subsection{CNN Architecture and Training}
\label{subsec:cnn_arch}

For each input representation, a CNN is trained for one-step-ahead regression using GAF images as inputs. The architecture consists of stacked convolutional layers with ReLU activations (Nair and Hinton) and max-pooling, followed by fully connected layers mapping to a scalar output. Training uses the same chronological 80--20 train/test split as the MLP experiments, with out-of-sample evaluation on the held-out 20\% test segment. A window length of $W=20$ observations is used across all series for comparability.

\subsection{Out-of-Sample Results and Interpretation}
\label{subsec:cnn_results}

Table~\ref{tab:cnn_gaf_comparison} reports out-of-sample prediction performance
for the CNN--GAF models over the three input representations.

\begin{table}[h!]
\centering
\caption{CNN--GAF out-of-sample prediction performance across three input
         representations (test set, 20\% holdout, horizon $h=1$).}
\label{tab:cnn_gaf_comparison}
\begin{tabular}{lrrr}
\toprule
\textbf{Input Series} & \textbf{RMSE} & \textbf{MAE} & \textbf{$R^2$} \\
\midrule
Levels (non-stationary, $I(1)$) & 5.823 & 4.829 & $-4.683$ \\
Log Returns (stationary, $I(0)$) & 0.004 & 0.003 & $-0.806$ \\
Frac.\ Diff.\ ($d=0.4$, near-stationary) & 0.871 & 0.750 & $-2.646$ \\
\bottomrule
\end{tabular}
\end{table}

A consistent pattern emerges across all three representations: every CNN--GAF
model gives a negative out-of-sample $R^2$, indicating performance inferior to
a naive mean predictor at every step. The levels model produces the largest
absolute errors (RMSE $= 5.82$, MAE $= 4.83$) and the most negative $R^2$
($= -4.68$), a direct consequence of feeding a non-stationary series into a model
trained on a different distributional regime than the one it encounters at test
time. Because GAF images are constructed from min-max-normalized windows, the
global rescaling is determined by local window extrema, which can shift
systematically as the level series trends through time; the CNN thus faces a
distribution shift between training and test images that is more severe than what
the MLP encounters with standardized lag vectors.

The log-return model achieves the smallest raw errors (RMSE $= 0.004$,
MAE $= 0.003$), reflecting the narrow range of return fluctuations. However, an
$R^2$ of $-0.81$ reveals that the model cannot beat the mean predictor,
indicating that the CNN model fails to extract any net predictive signal from the GAF
representation of return series. A plausible mechanism is that high-frequency
noise and heteroskedasticity, which dominate the return series, translate into
distinctive but unpredictable visual artifacts in the GAF images; the CNN may
fit these artifacts in-sample while generalizing poorly out-of-sample.

\begin{figure}[H]
  \centering
  \includegraphics[width=\linewidth]{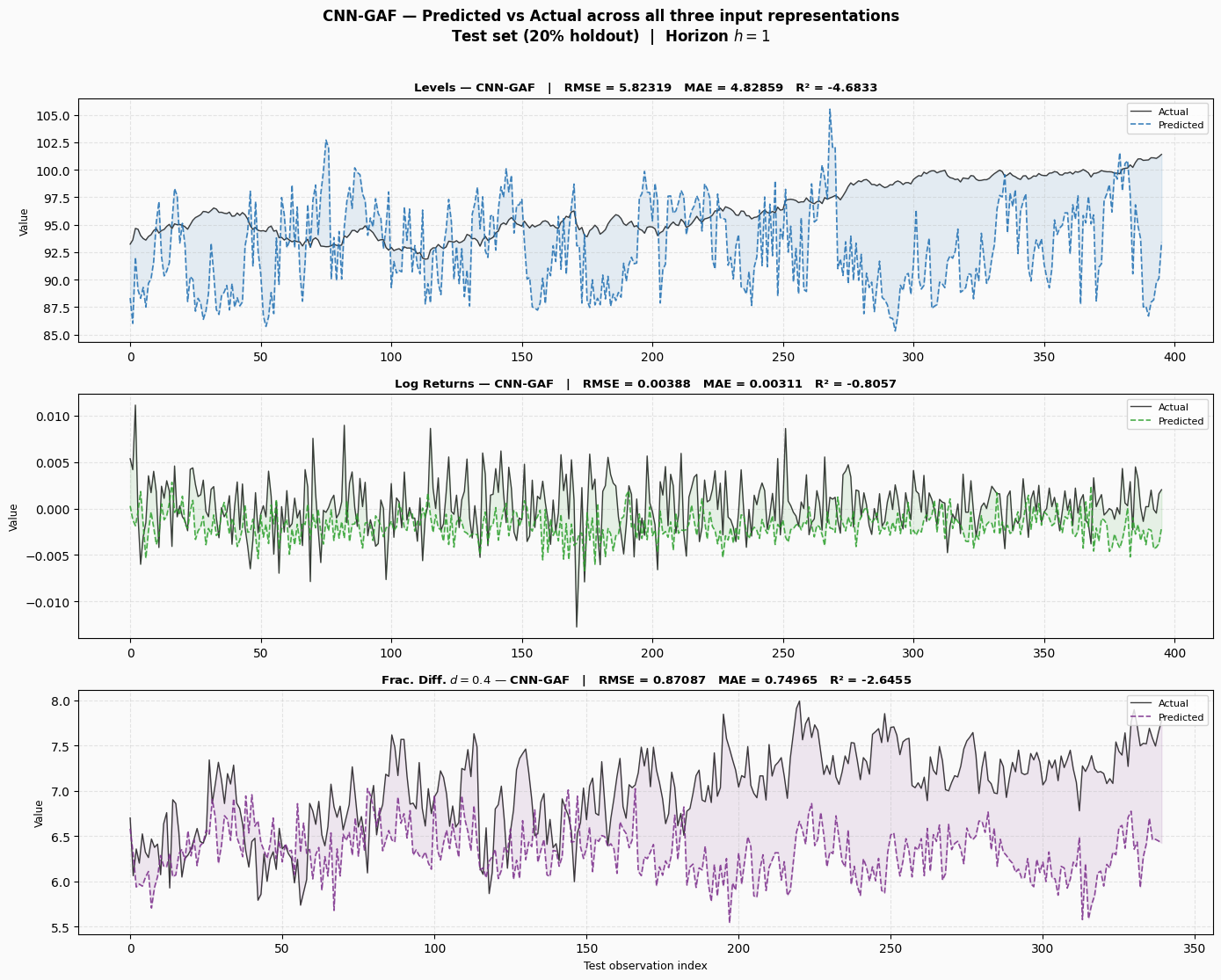}
  \caption{Evaluation of predictive performance for CNN--GAF models across the
           three input representations.}
  \label{fig:cnn_eval}
\end{figure}

The fractionally differenced model ($d=0.4$; RMSE $= 0.871$, $R^2 = -2.65$)
occupies an intermediate position in raw error magnitude, but its $R^2$ is more
negative than the log-return model. A likely explanation is that the $W=20$
window used to construct each GAF image truncates the very long-memory
information that fractional differencing was designed to preserve; the CNN model is
effectively asked to learn long-memory dynamics from a short-memory image,
creating an information bottleneck. Additionally, with a limited number of
training epochs and a relatively modest architecture, the CNN may fail to
identify the subtle linear mappings from image features to the next-step target,
or may overfit transient image artifacts that do not generalize.
Figure~\ref{fig:cnn_eval} summarizes the predictive performance diagnostics for
all three CNN--GAF models.

\section{Comparative Discussion: MLP versus CNN--GAF}
\label{sec:comparison}

\subsection{Inductive Bias and Representation}
\label{subsec:inductive_bias}

The sharp empirical divergence between MLP and CNN--GAF performance is best
understood through the lens of inductive bias (LeCun et al., ``Deep
Learning'') - the set of assumptions each architecture embeds about the structure
of the learning problem. The MLP receives the last $L$ values as an ordered
numerical vector and learns a nonlinear mapping
$\hat{y}_t = f_\theta(y_{t-1}, \ldots, y_{t-L})$ in which temporal ordering is
persisted explicitly by the feature construction. For level-like series, the MLP
can easily represent persistence-dominated dynamics by learning a near-identity
map on the first lag, directly approximating the Bayes-optimal forecast under
near-random-walk dynamics. The CNN model, in contrast, operates on a GAF image
$G_t \in \mathbb{R}^{W \times W}$ that emphasizes pairwise angular relationships
among points in a window, rather than the original numerical scale of the series.
This representation is powerful when the forecasting problem depends on shape,
motifs, or recurring local patterns (Wang and Oates; Fawaz et al.) - but it is
not naturally aligned with persistence-driven dynamics, where the most
informative single statistic for next-step prediction is the immediately preceding
value $y_{t-1}$.

For the $I(1)$ level series, the Bayes-optimal one-step predictor is close to
the identity $\hat{y}_t \approx y_{t-1}$, and the MLP achieves this by selecting
$L=1$ and learning the corresponding function. The CNN--GAF pipeline introduces
an intermediate representation that is not tailored to identity-like forecasting:
the GAF matrix may primarily encode the smoothness or curvature of the local
trajectory, discarding the simplest sufficient statistic for the one-step
prediction problem. For the return series, weak predictability under the
efficient market hypothesis (Fama) causes both architectures to collapse toward
trivial forecasts, but the MLP is less prone to fitting spurious visual artifacts
induced by the GAF transform, because MAE-minimizing training with early stopping
tends to shrink toward a conservative near-zero predictor. For the fractionally
differenced series, the residual persistence is naturally expressed in lag space,
and the MLP can approximate the implied smooth nonlinear corrections; the
CNN-GAF pipeline likely suffers from the combination of a short window
(information bottleneck) and sensitivity to transient image-level artifacts.

\subsection{Practical Implications}
\label{subsec:practical}

The results support a pragmatic conclusion: for one-step-ahead forecasting of
highly persistent financial levels and weakly autocorrelated returns, direct
lag-based models - including linear baselines and modest nonlinear MLPs - are
competitive and often difficult to improve upon. CNNs operating on GAF images may
be more appropriate when the forecasting problem is genuinely pattern-based - for
example, classification of market regimes, detection of recurring structural
motifs, or multi-step horizon prediction where the shape of recent dynamics
provides discriminative information (Fawaz et al.; Ismail Fawaz et al.). In the
present application, the non-stationarity of levels, the near-randomness of
returns under weak-form efficiency (Fama), and the long-memory nature of the
fractionally differenced series collectively favor architectures that preserve
the temporal ordering and numerical scale of lags. This reinforces the central
message of this paper: the choice of series transformation and the alignment
between architectural inductive bias and the statistical properties of the data
are often more consequential for predictive performance than architectural
complexity per se (Gu et al.; L\'{o}pez de Prado).

\newpage

\section{Conclusion}
\label{sec:conclusion}
This paper investigated the statistical properties and short-horizon predictability of the U.S.\ Aggregate Bond Index using modern deep learning methods, with emphasis on series transformation and architectural inductive bias. Index levels are found to be $I(1)$ processes with near-unit-root persistence and slowly decaying autocorrelations, while log returns are $I(0)$ with weak serial dependence, volatility clustering, and fat tails. Applying the fractional differencing procedure of L\'{o}pez de Prado yields an optimal order of approximately $d^* \approx 0.40$--$0.45$, producing a near-stationary representation that preserves low-frequency memory relative to full first differencing (Granger and Joyeux; Hosking).

Forecasting experiments show that MLP models trained on lagged vectors achieve near-baseline performance for levels (driven by persistence), near-zero performance for returns (consistent with weak-form efficiency (Fama)), and the strongest incremental performance for the fractionally differenced series, where moderate dependence remains. CNN--GAF models produce uniformly negative out-of-sample $R^2$ across all representations, suggesting a mismatch between the geometric image encoding and the persistence-dominated one-step forecasting task.

Overall, predictive performance is driven primarily by the transformation of the underlying series---its stationarity and memory structure---rather than model complexity. For practitioners in fixed-income machine learning, this highlights the importance of preprocessing and aligning model inductive bias with the statistical structure of the data before deploying sophisticated architectures. Future work may examine whether richer feature sets (cross-asset signals, macroeconomic indicators, or term-structure variables), alternative architectures such as LSTMs (Hochreiter and Schmidhuber) or attention-based transformers, or longer forecast horizons alter these conclusions, and whether CNN--GAF methods are better suited for classification or regime-detection tasks.

\newpage
\appendix
\section{Appendix: Fractional Differencing - Weights and Implementation}
\label{app:fracdiff}

The fractional differencing operator $(1-L)^d$ produces weights
\begin{equation}
w_k = (-1)^k \binom{d}{k}
    = (-1)^k \frac{d(d-1)\cdots(d-k+1)}{k!},
\end{equation}
with $w_0 = 1$, $w_1 = -d$, $w_2 = d(d-1)/2$, etc. For large $k$, the
weights decay at the hyperbolic rate $|w_k| \sim k^{-(1+d)}$, in contrast to
the finite-memory operator obtained at $d=1$, where all weights beyond $k=1$ are
exactly zero. This slow, hyperbolic decay is the defining property of long-memory
processes and distinguishes fractional differencing from exponential decay models
(Hosking; Granger and Joyeux). In practice, the infinite sum is truncated at a
finite lag $K$ determined by a weight-threshold criterion
($|w_K| < \epsilon$, typically $\epsilon = 10^{-5}$), a fixed window, or a
cumulative weight criterion. Truncation introduces a trade-off: larger $K$
preserves more long-memory structure but reduces the effective sample size to
$T_{\text{effective}} = T - K$ and increases computational cost; smaller $K$
improves efficiency but approximates the infinite operator more coarsely. When
$d=1$, the operator collapses to the standard first-difference operator $(1-L)$,
recovering $\tilde{P}_t = P_t - P_{t-1}$ with finite memory (Box et al.). For
$d \in (0,1)$, the operator applies a low-pass filter that attenuates stochastic
trend components while preserving long-memory dependence - the ``stationary but
maximally persistent'' transformation advocated by L\'{o}pez de Prado.

\newpage

\begin{workscited}

Baillie, Richard T. ``Long Memory Processes and Fractional Integration in
Econometrics.'' \textit{Journal of Econometrics}, vol.~73, no.~1, 1996,
pp.~5--59.

Bollerslev, Tim. ``Generalized Autoregressive Conditional Heteroskedasticity.''
\textit{Journal of Econometrics}, vol.~31, no.~3, 1986, pp.~307--327.

Box, George E. P., et al. \textit{Time Series Analysis: Forecasting and Control}.
5th ed., Wiley, 2015.

Campbell, John Y., and Robert J. Shiller. \textit{The Econometrics of Financial
Markets}. Princeton University Press, 2012.

Cont, Rama. ``Empirical Properties of Asset Returns: Stylized Facts and
Statistical Issues.'' \textit{Quantitative Finance}, vol.~1, no.~2, 2001,
pp.~223--236.

Dickey, David A., and Wayne A. Fuller. ``Distribution of the Estimators for
Autoregressive Time Series with a Unit Root.'' \textit{Journal of the American
Statistical Association}, vol.~74, no.~366, 1979, pp.~427--431.

Engle, Robert F. ``Autoregressive Conditional Heteroscedasticity with Estimates
of the Variance of United Kingdom Inflation.'' \textit{Econometrica}, vol.~50,
no.~4, 1982, pp.~987--1007.

Engle, Robert F., and Clive W. J. Granger. ``Co-Integration and Error
Correction: Representation, Estimation, and Testing.''
\textit{Econometrica}, vol.~55, no.~2, 1987, pp.~251--276.

Fama, Eugene F. ``Efficient Capital Markets: A Review of Theory and Empirical
Work.'' \textit{Journal of Finance}, vol.~25, no.~2, 1970, pp.~383--417.

Fawaz, Hassan Ismail, et al. ``Deep Learning for Time Series Classification:
A Review.'' \textit{Data Mining and Knowledge Discovery}, vol.~33, no.~4, 2019,
pp.~917--963.

Granger, Clive W. J., and Roselyne Joyeux. ``An Introduction to Long-Memory
Time Series Models and Fractional Differencing.'' \textit{Journal of Time Series
Analysis}, vol.~1, no.~1, 1980, pp.~15--29.

Gu, Shihao, et al. ``Empirical Asset Pricing via Machine Learning.''
\textit{Review of Financial Studies}, vol.~33, no.~5, 2020, pp.~2223--2273.

Hamilton, James D. \textit{Time Series Analysis}. Princeton University Press,
1994.

Hochreiter, Sepp, and J\"{u}rgen Schmidhuber. ``Long Short-Term Memory.''
\textit{Neural Computation}, vol.~9, no.~8, 1997, pp.~1735--1780.

Hornik, Kurt, et al. ``Multilayer Feedforward Networks Are Universal
Approximators.'' \textit{Neural Networks}, vol.~2, no.~5, 1989, pp.~359--366.

Hosking, J. R. M. ``Fractional Differencing.'' \textit{Biometrika}, vol.~68,
no.~1, 1981, pp.~165--176.

Ioffe, Sergey, and Christian Szegedy. ``Batch Normalization: Accelerating Deep
Network Training by Reducing Internal Covariate Shift.'' \textit{Proceedings of
the 32nd International Conference on Machine Learning}, 2015, pp.~448--456.

Ismail Fawaz, Hassan, et al. ``InceptionTime: Finding AlexNet for Time Series
Classification.'' \textit{Data Mining and Knowledge Discovery}, vol.~34, no.~6,
2019, pp.~1936--1962.

Kingma, Diederik P., and Jimmy Ba. ``Adam: A Method for Stochastic
Optimization.'' \textit{arXiv}, 2014, arXiv:1412.6980.

LeCun, Yann, et al. ``Backpropagation Applied to Handwritten Zip Code
Recognition.'' \textit{Neural Computation}, vol.~1, no.~4, 1989, pp.~541--551.

---. ``Deep Learning.'' \textit{Nature}, vol.~521, no.~7553, 2015, pp.~436--444.

L\'{o}pez de Prado, Marcos. \textit{Advances in Financial Machine Learning}.
Wiley, 2018.

Mandelbrot, Benoit B. ``The Variation of Certain Speculative Prices.''
\textit{Journal of Business}, vol.~36, no.~4, 1963, pp.~394--419.

Meese, Richard A., and Kenneth Rogoff. ``Empirical Exchange Rate Models of the
Seventies: Do They Fit Out of Sample?'' \textit{Journal of International
Economics}, vol.~14, no.~1--2, 1983, pp.~3--24.

Nair, Vinod, and Geoffrey E. Hinton. ``Rectified Linear Units Improve Restricted
Boltzmann Machines.'' \textit{Proceedings of the 27th International Conference
on Machine Learning}, 2010, pp.~807--814.

Nelson, Charles R., and Charles R. Plosser. ``Trends and Random Walks in
Macroeconomic Time Series: Some Evidence and Implications.''
\textit{Journal of Monetary Economics}, vol.~10, no.~2, 1982, pp.~139--162.

Rumelhart, David E., et al. ``Learning Representations by Back-Propagating
Errors.'' \textit{Nature}, vol.~323, no.~6088, 1986, pp.~533--536.

Said, Said E., and David A. Dickey. ``Testing for Unit Roots in
Autoregressive-Moving Average Models of Unknown Order.''
\textit{Biometrika}, vol.~71, no.~3, 1984, pp.~599--607.

Srivastava, Nitish, et al. ``Dropout: A Simple Way to Prevent Neural Networks
from Overfitting.'' \textit{Journal of Machine Learning Research}, vol.~15,
no.~1, 2014, pp.~1929--1958.

Wang, Zhiguang, and Tim Oates. ``Imaging Time-Series to Improve Classification
and Imputation.'' \textit{Proceedings of the 24th International Joint Conference
on Artificial Intelligence}, 2015, pp.~3939--3945.

\end{workscited}

\end{document}